\documentclass[conference]{IEEEtran}
\IEEEoverridecommandlockouts
\usepackage{cite}
\usepackage{amsmath,amssymb,amsfonts}
\usepackage{algorithmic}
\usepackage{graphicx}
\usepackage{textcomp}
\usepackage{xcolor}
\usepackage{acronym}
\usepackage{siunitx}
\usepackage{multirow}
\usepackage{placeins}
\usepackage{tikz}
\usepackage{pgfplots}
\pgfplotsset{compat=1.18}

\def\BibTeX{{\rm B\kern-.05em{\sc i\kern-.025em b}\kerso n-.08em
    T\kern-.1667em\lower.7ex\hbox{E}\kern-.125emX}}
\begin{document}

\usetikzlibrary{positioning, fit, backgrounds}

\newcommand{\com}[1]{\textcolor{red}{#1}}

\title{Throughput Requirements for RAN Functional Splits in 3D Networks\\
{\footnotesize \textsuperscript{}}
\thanks{This research was supported in part by the German Federal Ministry of Education and Research (BMBF) within the projects Open6GHub under grant number 16KISK016 and 6G-TakeOff under grant number 16KISK068 as well as the European Space Agency (ESA) under contract number 4000139559/22/UK/AL (AIComS).}
}
\author{\IEEEauthorblockN{MohammadAmin Vakilifard, Tim Düe, Mohammad Rihan, Maik Röper,\\ Dirk Wübben, Carsten Bockelmann, Armin Dekorsy}
\IEEEauthorblockA{\textit{Department of Communications Engineering},
\textit{University of Bremen},
Bremen, Germany. \\
\{vakilifard, duee, elmeligy, roeper, wuebben, bockelmann, dekorsy\}@ant.uni-bremen.de}\\}

\DeclareSIUnit{\sample}{S}
\newacro{RU}{Radio Unit}
\newacro{DU}{Distributed Unit}
\newacro{CU}{Center Unit}
\newacro{NTN}{Non-Terrestrial Network}
\newacroplural{NTN}[NTNs]{Non-Terrestrial Networks}
\newacro{FS}{Functional Split}
\newacro{gNB}{gNodeB}
\newacroplural{gNB}[gNBs]{gNodeBs}
\newacro{LEO}{Low Earth Orbit}
\newacroplural{LEO}[LEO]{Low Earth Orbits}
\newacro{UAV}{Unmanned Aerial Vehicle}
\newacro{UL}{Uplink}
\newacro{UE}{User Equipment}
\newacro{5GC}{5G Core}
\newacro{FFT}{Fast Fourier Transform}
\newacro{IFFT}{Inverse Fast Fourier Transform}
\newacro{mMTC}{Massive Machine Type Communication}
\newacro{eMBB}{Enhanced Mobile Broadband}
\newacro{NR}{New Radio}
\newacro{O-RAN}{Open-RAN}
\newacro{DL}{Downlink}
\newacro{FH}{Fronthaul}
\newacro{FDD}{Frequency Division Duplexing}
\newacro{BW}{Bandwidth}
\newacro{FH-BW}{Fronthaul-\ac{BW}}
\newacro{CSI}{Channel State Information}
\newacro{RB}{Resource Block}
\newacroplural{RB}[RBs]{Resource Blocks}
\newacro{CP}{Cyclic Prefix}
\newacro{IBM}{Information Bottleneck Method}
\newacro{HAPS}{High Altitude Platform Station}
\newacro{HIBS}{High Altitude Platforms as IMT base Station}
\newacro{UAV}{Unmanned Aerial Vehicle}
\newacroplural{HAPS}[HAPS]{High Altitude Platform Station}
\newacro{LLR}{Log-Likelyhood Ratio}
\newacroplural{LLR}[LLRs]{Log-Likelihood Ratios}
\newacro{LDPC}{Low density parity check}
\newacro{BER}{Bit Error Rate}
\newacro{RE}{Resource Element}
\newacroplural{RE}[REs]{Resource Elements}
\newacro{CDL}{Cluster Delay Line}
\newacro{MCS}{modulation and coding scheme}
\newacro{ISL}{Inter Satellite Link}
\newacro{FEC}{Forward Error Correction}
\newacro{mMTC}{massive machine-type communications}
\newacro{RLC}{Radio Link Control}
\newacro{SINR}{Signal to Interference and Noise ratio}
\newacro{RAN}{Radio Access Network}
\maketitle

\begin{abstract}
The rapid growth of non-terrestrial communication necessitates its integration with existing terrestrial networks, as highlighted in 3GPP Releases 16 and 17. This paper analyses the concept of functional splits in 3D networks. To manage this complex structure effectively, the adoption of a Radio Access Network (RAN) architecture with \ac{FS} offers advantages in flexibility, scalability, and cost-efficiency. RAN achieves this by disaggregating functionalities into three separate units. Analogous to the terrestrial network approach, 3GPP is extending this concept to non-terrestrial platforms as well. This work presents a general analysis of the requested \ac{FH} data rate on the feeder link between a non-terrestrial platform and the ground-station in the \ac{UL} for two typical \ac{NTN} scenarios, \ac{LEO} satellite and \ac{HAPS}. One of the trade-offs for each split option is  between the \ac{FH} data rate and the resulting complexity based on the use case scenario. Since non-terrestrial nodes face more limitations regarding power consumption and complexity onboard in comparison to terrestrial ones, the throughput requirements and suitability of different functional split options must be investigated for this use case as well. 
\end{abstract}

\begin{IEEEkeywords}
Functional Split (FS), 3D Networks, Non-Terrestrial Communication, 3GPP, Fronthaul Throughput, NTN
\end{IEEEkeywords}

\section{Introduction}\label{ch: Introduction}
Current terrestrial networks face limitations in achieving complete coverage due to restricted coverage areas and geographical barriers. Network outages during natural disasters can further disrupt critical operations, jeopardizing lives and property \cite{b1, b2}. Three-dimensional (3D) wireless communication networks represent the next frontier in network deployment. These networks leverage a layered approach, integrating space-, air-, and terrestrial-based communication links to achieve seamless and ubiquitous coverage. As we approach the 2030s, the anticipated roll-out of 6G, full integration of non-terrestrial communication links becomes crucial \cite{b3}. This includes satellites, mega-constellations in \acp{LEO}, and airborne platforms such as \acf{HAPS} and \acp{UAV} working in tandem with terrestrial networks. The high number of potential \ac{gNB} nodes due to non-terrestrial integration necessitates flexibility, robustness, cost-effectiveness, and energy efficiency in these new platforms. Consequently, integrating the concept of \acf{FS} with non-terrestrial platforms becomes a critical step towards achieving these goals.
\begin{figure}[bpt]
\centering
\includegraphics[width=1.0\linewidth]{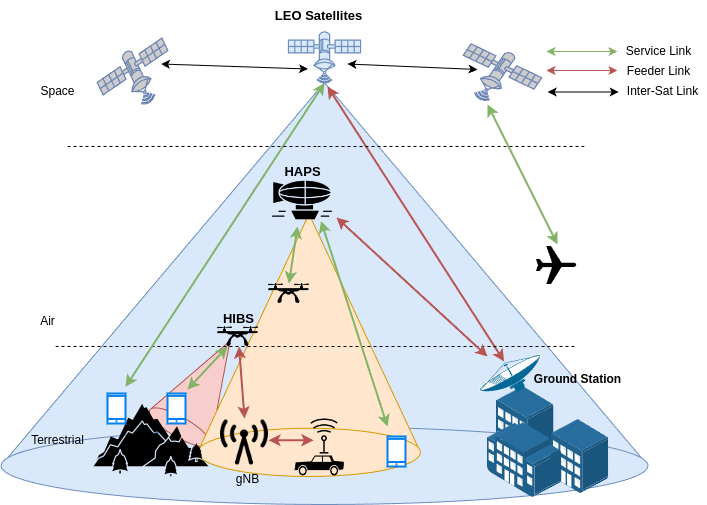}
\caption{3D Network architecture, composed of \ac{LEO} satellites, \ac{HAPS} and \ac{UAV} as two form of \ac{HIBS} with the different beam size of each Non terrestrial node on the ground.}
\label{fig:3d_network}
\end{figure}
\label{ch:FS_for_3D}

The 3GPP Release 15 introduced the adaptable design for the 5G \ac{RAN} \cite{b4}. This design disaggregates the traditional \ac{gNB} into three logical nodes: the \acf{RU}, \acf{DU}, and \acf{CU}. So far, 3GPP has introduced in principle eight \ac{FS} options for \ac{RAN} in terrestrial networks. In the recent years, the O-RAN alliance, another major player, has focused on physical layer splits and presented its own split based architecture of \ac{RAN} \cite{b5}. 

Different \ac{FS} options necessitate varying data rates to transmit the same information between nodes as calculated in \cite{b6}. These requirements depend on the deployment scenario and the characteristics of the interface between the split components. 3GPP has addressed the calculation of \ac{FH} data rate for LTE and \ac{NR} in \cite{b7, b8, b9} in their respective specifications.
So far, most investigations on \ac{FS} are performed for terrestrial networks \cite{b10, b11}, while some papers such as \cite{b28} and \cite{b29} has analyzed the feasibility of RAN functional split in \ac{NTN} but by considering \ac{FH} data rate based on the terrestrial network. Therefore specific constraints in \ac{NTN} scenario are not completely investigated.

This paper proposes a general formulation to determine the \ac{FH} data rate needed for implementing \ac{FS} in the \acf{UL} with non-terrestrial platforms. The \ac{FH} data rate describes the amount of data needed to be transmitted between a terrestrial and non-terrestrial node during on the feeder link. Additionally, we suggest which of the investigated physical layer split options is most suitable for non-terrestrial nodes.

\section{Functional Split for 3D Networks}
\subsection{3D Networks}
3D Networks extend conventional terrestrial networks with air- and space-borne network nodes, as depicted in Fig.~\ref{fig:3d_network}.
The \ac{NTN} nodes, i.e., the air- and spaceborne network nodes such as \ac{LEO} satellites and \ac{HIBS}, provide connectivity between a terrestrial \ac{UE} and a ground station, which is connected to the Core Network.

Accordingly, we distinguish between two communication links:
\begin{enumerate}
\item Service Link: The communication link connecting \ac{UE} and the non-terrestrial platform, which hosts some part of RAN functionalities.
\item Feeder Link: The communication link between the non-terrestrial platform and the ground which host the rest of RAN functionalities and is connected to the Core Network.
\end{enumerate}

\subsection{Functional Split in Terrestrial Network}
\begin{figure}[bpt]
    \centering
    \includegraphics[width=1.0\linewidth]{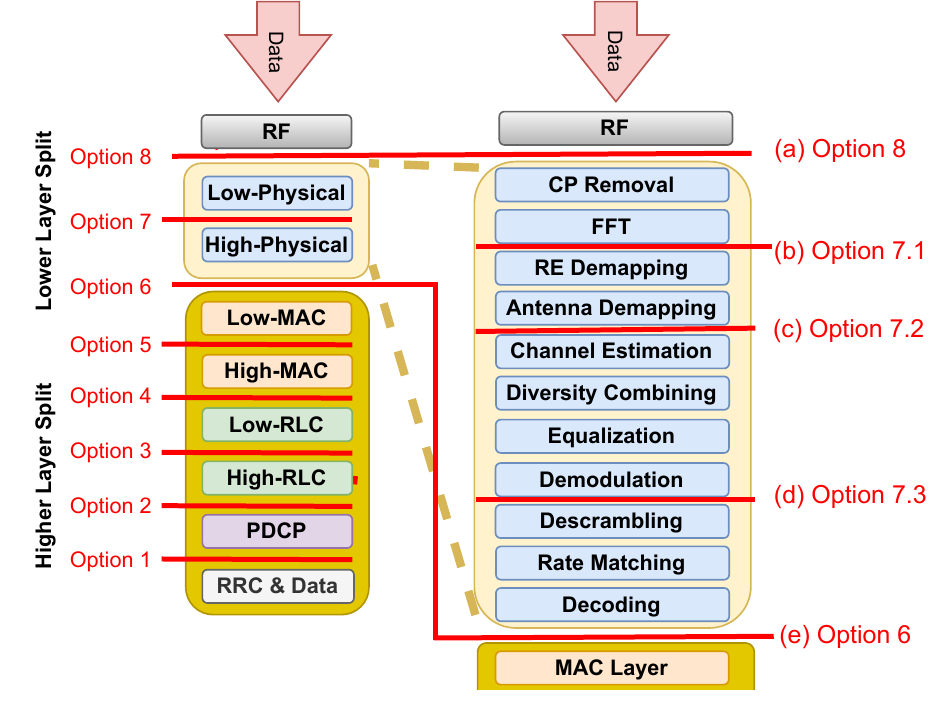}
    \caption{3GPP \ac{FS} in \ac{UL}. On the left there are eight \ac{FS} options introduced by 3GPP, and on the right there are lower layer split options in baseband processing unit.}
    \label{fig:low_split}
\end{figure}

3GPP introduced in principle eight \ac{FS} options for a \ac{gNB}, as depicted in Fig.~\ref{fig:low_split}, whereas options 4 and 5 are deemed impractical for implementation \cite{b5}. O-RAN also has introduced architectures, which decompose network functions into distinct components that can be implemented on various hardware, which have garnered increased attention \cite{b14}. O-RAN primarily concentrates on option 7 of the \ac{FS} proposed by 3GPP, with a focus on separating the \ac{RU} and \ac{DU} functions \cite{b15}. Depending on specific \acp{FS}, there is a trade-off between computational complexity onboard of the platform and data rate demand on interface link. Other trade-offs, such as timing requirements, latency restrictions, and link performance, also are of high importance but not in the context of this paper.

\subsection{Functional Split for Non-Terrestrial Network}
3GPP has introduced three non-terrestrial access architectures for satellites, including the transparent payload-based, where the satellite acts as a relay between the user and the Core \cite{b16}; a regenerative payload-based, where the \ac{gNB} is fully implemented on the satellite board \cite{b5}; and a regenerative satellite-based \ac{gNB}-\ac{DU}, which considers split option 2 with the complete \ac{RU} and \ac{DU} implemented onboard the satellite \cite{b17}.
O-RAN has not yet officially considered its architecture for \ac{NTN}, while in \cite{b18} O-RAN based \ac{NTN} has been investigated.

While the 3GPP has currently designated option 2 for \acp{NTN}, the push towards a unified 3D network with diverse communication elements necessitates exploring lower layer split options \cite{b19, b20} to reduce complexity by not realizing full \ac{gNB} on the non-terrestrial nodes. For \ac{NTN}, such lower layer splits lead to a reduced complexity of the \ac{NTN} nodes. 

The reasons why 3GPP has so far not favored lower layer splits for \ac{NTN} are timing issues and latency. In terrestrial networks, the necessity to quickly adjust to changes in \ac{CSI} drives the demand for reduced latency in lower layer splits, as e.g. beamforming requires precise \ac{CSI}.

However, if the position of the ground station node and user nodes are known sufficiently well to the non-terrestrial node, the requirements for precise \ac{CSI} reduce significantly. As the beamforming on space-born platforms mainly sets the beam on specific positions on the Earths surface and does not adjust to fast changing channel conditions, the positional information can be used to infer the \ac{CSI} information as shown in \cite{b21}, relaxing the latency requirements on the feeder link, which results in make the \ac{FS} possible. 



The main challenges regarding \ac{FS} for non-terrestrial platforms can be listed as:
\begin{enumerate}
    \item Timing requirements fulfilling 3GPP constraints. 
    \item Level of computational complexity needed onboard, which correlates with energy consumption and cost;
    \item High \ac{FH} data rate over the feeder-link;
    \item Constrained memory capacity requirements onboard.
\end{enumerate}

As seen in Fig.~\ref{fig:low_split}, split options are classified into high layer and low layer splits. High layer splits enable effective data storage on non-terrestrial systems by reducing memory requirements as a result of more processing operations. High layer \ac{FS} reduce the needed \ac{FH} data rate and need higher timing requirements. However, they also raise the complexity and power consumption of the baseband processing.
Lower \ac{FS}, on the other hand, may be advantageous for non-terrestrial nodes since they reduce hardware complexity, as this study explores with options 8, 7, and 6. 
The necessity for a balanced approach in the design and implementation of access architectures for non-terrestrial platforms is highlighted by this trade-off.

\section{System Model}

\label{ch:model_and_formulation}
By considering the regenerative access architecture as baseline and prior works of \cite{b22}, \cite{b23} and \cite{b24}, we focus on the \acf{UL} case from a \ac{UE} to the Core Network through a non-terrestrial node. Fig.~\ref{fig:low_split} shows the block diagram of the baseband processing chain for each \ac{FS} option from 8 to 6 and their involving functionalities. By moving from \ac{FS} option 8 towards 6, the \ac{FH} data rate and memory requirements are reduced, but the computational complexity and power consumption will increase.

We consider a non-terrestrial communication node operational at the height of $h_{\mathrm{C}}$, so that the number of cells on the ground it can cover is $N_{\mathrm{C}}$. Each cell is covered by $N_{\mathrm{B}}$ beams. Each beam is created by $L$ number of antenna elements. Therefore, the total number of active antennas is $N_{\mathrm{ant}} = N_{\mathrm{C}} \cdot N_{\mathrm{B}} \cdot L$. 
We perform our analysis for two scenarios: 
\begin{itemize}
    \item Scenario 1 (SC1): A \ac{LEO} satellite at the height $h_{\mathrm{C}}$ = \SI{600}{\km} covers $N_{\mathrm{C}} = 19$ cells based on \cite{b16}
    \item Scenario 2 (SC2): A \ac{HIBS} at the height $h_{\mathrm{C}}$ = \SI{10}{\km} covering $N_{\mathrm{C}} = 8$ cells based on \cite{b22}.
\end{itemize} 
Without loss of generality and for the sake of comparison, we consider based on \cite{b4}, \cite{b16} and \cite{b25}  that both \ac{LEO} satellite and the \ac{HIBS} use the same array of antennas and one beam per cell is considered ($N_{\mathrm{B}} = 1$). Each beam can support a bandwidth of $B_{\mathrm{BW}}$. The total number of covered UEs in each beam is:
\begin{equation}
    N_{\mathrm{UE}} = A_{\mathrm{Beam}} \cdot \rho_{\mathrm{UE}}
\end{equation}
where:
\begin{itemize}
    \item $A_{\mathrm{Beam}}$ is the beam foot print size on ground equal to $\pi \cdot r_{\mathrm{Beam}}^{2}$ which $r_{\mathrm{Beam}}$ is the beam radius as function of platform height, its equivalent antenna aperture and carrier frequency band. Therefore we have $r_{\mathrm{Beam}_{\mathrm{SC1}}}$ equal to \SI{25}{\km} for S band and \SI{10}{\km} for Ka band.
    \item $\rho_{\mathrm{UE}}$ is UEs density per square kilometer $\frac{\mathrm{UEs}}{\mathrm{km^{2}}}$ , which varies depending on the type of communication such as \ac{eMBB} or \ac{mMTC}.
\end{itemize}
Since different altitudes $h_{\mathrm{C}}$ and corresponding values of $N_{\mathrm{UE}}$ result in different \ac{SINR} and consecutively different Modulation and Coding Scheme, but here in the rest of the paper without loss of generality and sake of simplicity, it is assumed that all covered UEs use the same modulation order $M$ and code rate $R_{\mathrm{c}}$ for both scenarios. 

Based on \cite{b7} upon reception, each sample after Analog-to-Digital converter is represented by $Q_\mathrm{T}$ bits. The service link data rate is calculated by using
\begin{equation}
\label{eq:eq1}
    R_{\text{Service Link}} = N_{\mathrm{C}} \cdot N_{\mathrm{B}}\cdot N_{UE}\cdot B_{\mathrm{BW}} \cdot L \cdot S\;,
\end{equation}

in which $S$ is the over sampling rate. All the example parameters are reported with details in Table~\ref{tab:tab1}. The values are chosen based on different scenarios in the standard 3GPP TR38.821 \cite{b16}, \cite{b22} and \cite{b25}.

\begin{table}[btp]
\centering
\caption{Parameters for \ac{FH} calculations}
\label{tab:tab1}
\resizebox{\columnwidth}{!}{%
\fontsize{14}{16}\selectfont
\begin{tabular}{|c|c|cccc|}
\hline
Variable Name &
  Meaning &
  \multicolumn{4}{c|}{Value} \\ \hline
$N_{\mathrm{C}}$ &
  No. of covered cells &
  \multicolumn{2}{c|}{SC1: 19} &
  \multicolumn{2}{c|}{SC2: 8} \\ \hline
$B_{\mathrm{BW}}$ &
  \ac{BW} per beam &
  \multicolumn{2}{c|}{S band: \SI{30}{\mega \hertz}} &
  \multicolumn{2}{c|}{Ka band: \SI{400}{\mega \hertz}} \\ \hline
\multirow{2}{*}{$r_{\mathrm{Beam}}$} &
  \multirow{2}{*}{Beam radius on the ground} &
  \multicolumn{2}{c|}{SC1} &
  \multicolumn{2}{c|}{SC2} \\ \cline{3-6} 
 &
   &
  \multicolumn{1}{c|}{S:\SI{25}{\km}} &
  \multicolumn{1}{c|}{Ka:\SI{10}{\km}} &
  \multicolumn{1}{c|}{S: \SI{6}{\km}} &
  Ka:\SI{6}{\km} \\ \hline
$\rho_{\mathrm{UE}}$ &
  UEs density &
  \multicolumn{2}{c|}{\ac{eMBB}: 0.1 $\frac{\mathrm{UEs}}{\mathrm{km^{2}}}$} &
  \multicolumn{2}{c|}{\ac{mMTC}: 500 $\frac{\mathrm{UEs}}{\mathrm{km^{2}}}$} \\ \hline
\multirow{2}{*}{$PR$} &
  \multirow{2}{*}{Reference Peak Rate} &
  \multicolumn{2}{c|}{eMBB} &
  \multicolumn{2}{c|}{mMTC} \\ \cline{3-6} 
 &
   &
  \multicolumn{1}{c|}{S: \SI{2}{\mega \bit / \second}} &
  \multicolumn{1}{c|}{Ka: \SI{100}{\mega \bit / \second}} &
  \multicolumn{1}{c|}{S: \SI{0.256}{\mega \bit / \second}} &
  Ka:\SI{7}{\mega \bit / \second} \\ \hline
$BW_{\mathrm{ref}}$ &
  Reference bandwidth &
  \multicolumn{2}{c|}{S: \SI{5}{\mega \hertz}} &
  \multicolumn{2}{c|}{Ka: \SI{100}{\mega \hertz}} \\ \hline
$N_{\mathrm{B}}$ &
  No. of beams per cell &
  \multicolumn{4}{c|}{1} \\ 
  \hline
  $L$ &
  No.antenna elements create a beam &
  \multicolumn{4}{c|}{2} \\ 
  \hline
  $S$ &
  Over sampling rate &
  \multicolumn{4}{c|}{\SI{1}{\sample / \second}} \\ 
  \hline
  $Q_{\mathrm{T}}$  &
  Quantization in time &
  \multicolumn{4}{c|}{\SI{16}{\bit}s} \\ 
  \hline
  $Q_{\mathrm{F}}$ &
  Quantization in frequency &
  \multicolumn{4}{c|}{\SI{10}{\bit}s} \\ 
  \hline
$\tau_{\mathrm{CP}}$    & CP duration &
  \multicolumn{4}{c|}{4.688 \si{\micro \second}} \\ 
  \hline
$\tau_{\mathrm{subframe}}$   & OFDM subframe duration  & \multicolumn{4}{c|}{\SI{1}{\milli \second}} \\ 
\hline
$\mu$ & OFDM numerology & \multicolumn{4}{c|}{0} 
\\ \hline
$\eta$           & Utilization factor    & \multicolumn{4}{c|}{0.6}       
\\ \hline
$N_{\mathrm{S}}$ & No. symbols & \multicolumn{4}{c|}{14}
\\ \hline
$N_\mathrm{{SC}}$ & No. of subcarriers & \multicolumn{4}{c|}{12}
\\ \hline
$N_{\mathrm{Data}}$ & No. of data REs  & \multicolumn{4}{c|}{110}  
\\ \hline
$M$       & No. of symbols per beam   & \multicolumn{4}{c|}{2~:~256 QAM}  
\\ \hline
$Q_{\mathrm{LLR}}$       & Quantization per LLR & \multicolumn{4}{c|}{\SI{3}{\bit}s}          
\\ \hline
$R_{\mathrm{c}}$             & Code rate   & \multicolumn{4}{c|}{0.64 for $M = 16$}           
\\ \hline
$N_{\mathrm{L}}$     & No. Layer    & \multicolumn{4}{c|}{8}  \\ \hline
$M_{\mathrm{ref}}$     & Reference modulation order    & \multicolumn{4}{c|}{2 for $M = $ 2 and 4, 4 for $M = 16$, 16 for $M = 64$, 64 for $M = 256$}
\\ \hline
$p_{\mathrm{UL}}$     & Fraction of UEs requesting \ac{UL}& \multicolumn{4}{c|}{1 for a fully loaded scenario; 0 for an unloaded system}
\\ \hline
$S_{\mathrm{UL}}$     & Average content size in \ac{UL} by UEs    & \multicolumn{4}{c|}{\SI{30}{\byte} based on \cite{b8}}
\\ \hline
  
\end{tabular}%
}
\end{table}

\section{Fronthaul Rate Calculation for the Functional Split Options}
\label{ch:FH_BW_calc}
In this section, we utilize the general formulation and variable definitions provided in Section~\ref{ch:model_and_formulation} to derive the calculation of the \ac{FH} data rates of the considered \ac{FS} options by 3GPP as shown in Fig.~\ref{fig:low_split}. Additionally, we discuss their respective advantages and disadvantages. 

\subsection{Option 8: Time Domain}
In this \ac{FS} option, only RF-related processes are performed on the non-terrestrial node. These include analog-to-digital conversion and quantization at each antenna element's output. Assuming perfect sampling, no oversampling occurs, so data volume doesn't increase. The received signal is quantized in the time domain using $Q_{\mathrm{T}}$ \SI{}{\bit}s per sample per axis (In-Phase and Quadrature), resulting in each sample being represented by $2 \cdot Q_{\mathrm{T}}$ \SI{}{\bit}. Considering the maximum transmission bandwidth in $R_{\mathrm{ServiceLink}}$ from cells to the non-terrestrial platform we can write:
\begin{equation}
    R_{\mathrm{Opt8}} =  R_{\text{Service Link}} \cdot 2 \cdot Q_\mathrm{T} \;.
\end{equation}

The data to be transmitted over the \ac{FH} are I/Q samples, which in case of even no user data (no load) still exists. The primary advantage of option 8 lies in its low onboard complexity leading to reduced power consumption attributed to minimal signal processing. However, a notable drawback is the very high \ac{FH} data rate on the feeder link, as $Q_{\mathrm{T}}$ (\SI{16}{\bit}s is a typical value) is relatively high for time-domain signal to a high represent the signals adequately. 

\subsection{Option 7.1: Frequency Domain w/o CP}
This option involves the removal of symbols designated for OFDM \ac{CP}. The \ac{CP} length is a process dependent variable, represented by the OFDM numerology $\mu$ for each subframe. The duration of a single subframe $\tau_{\mathrm{subframe}}$ for \ac{NR} is standardized to \SI{1}{\milli \second} \cite{b5}.
The total duration of the CPs per frame is $N_\mathrm{S}\cdot (\mu + 1)\cdot \tau_{\mathrm{CP}}$.
After the \ac{CP} removal, the signal is transformed to the frequency domain using a \ac{FFT}. We can quantize level of the symbols in the frequency domain as $Q_\mathrm{F}$. $Q_{\mathrm{T}}$ and $Q_{\mathrm{F}}$ denote the number Q-levels if messages are transmitted over the \ac{FH}. In frequency domain less bits are sufficient therefore $Q_\mathrm{F} < Q_{\mathrm{T}}$ and typically is $Q_\mathrm{F} = \SI{10}{\bit}s$. The rate for this option can be expressed as

\begin{equation}
    R_{\mathrm{Opt7.1}} = R_{\mathrm{Opt8}} \cdot \frac{\tau_{\mathrm{subframe}}}{\tau_{\mathrm{subframe}} + N_{\mathrm{S}} \cdot(\mu + 1) \cdot \tau_{\mathrm{CP}}} \cdot \frac{Q_{\mathrm{F}}}{Q_{\mathrm{T}}}\;.
\end{equation}

This \ac{FS} results in the reduction of data rate, as both terms are less than $1$. Option 7.1 does add complexity onboard, however, it is still relatively low, as the required operations are basic and efficient hardware solutions exist. 

\subsection{Option 7.2: Resource-Element Demapping}
This option follows a procedure outlined in \cite{b26}, where the maximum number of layers is $N_{\mathrm{L}} = 8$ which is the same for \ac{NTN} as well. \ac{RE} demapping eliminates unused resource elements in the received signals from users which is determined by the system's actual load, referred to as utilization factor $\eta$, which includes antenna logical ports demapping according to \cite{b27}. The data rate of option 7.2 can be written:

\begin{equation}
    R_{\mathrm{Opt7.2}} = R_{\mathrm{Opt7.1}} \cdot \eta \;.
\end{equation}

 The inclusion of $\eta$ leads to a reduction in the \ac{FH} rate, as we now forward only the used \acp{RE}, therefore this option depends on the actual load of the system. This \ac{FS} option is very significant in scenarios with very low utilization, expressed by $\eta \ll 1$. The value of $\eta$ highly depends on the scenario. This \ac{FS} is currently favored by 3GPP for terrestrial networks and could be an effective option for non-terrestrial platforms as well.
 
\subsection{Option 7.3: Equalization and Demodulation}
This option introduces a split in \ac{UL} motivated by \ac{FS} option 7.3 for terrestrial networks of the 3GPP. This \ac{FS} option includes all the functionalities requiring \ac{CSI}, which are channel estimation, diversity combining, equalization, and demodulation. Therefore, option 7.3 omits the necessity to forward \ac{CSI} to the ground station.

Here, the pilots and other reference symbols like Sounding Reference Signals (SRS) and Phase-tracking Reference Signals (PTRS), employed to gain \ac{CSI}, can be removed, decreasing the \ac{FH} data rate. Equalization is performed for each beam, which reduces the throughput by a factor of $\frac{1}{L}$ after combining multiple antenna streams into a single beam. After demodulation, there are $\log_{2}M$ code bits per symbol, each \SI{}{\bit} is expressed by \ac{LLR} with $Q_{\mathrm{LLR}}$ \SI{}{\bit}.
The resulting \ac{FH} data rate is given by
\begin{equation}
\label{eq:fs_7_3}
    R_\mathrm{{Opt7.3}} = R_{\mathrm{Opt7.2}} \cdot \frac{N_{\mathrm{Data}}}{N_{\mathrm{S}} \cdot N_{\mathrm{SC}}} \cdot \frac{1}{L} \cdot \frac{\log_{2}M}{Q_\mathrm{F}} \cdot Q_{\mathrm{LLR}} \;.
\end{equation}
 
Here, $N_{\mathrm{Data}}$ represents all \ac{RE}s dedicated solely for data transmission, whereas   $N_\mathrm{S}\cdot N_\mathrm{SC}$ represents the total number of \acp{RE} after the \ac{CP} removal.

By increasing modulation order $M$, the \ac{FH} data rate increases as well, as $Q_{\mathrm{LLR}}$ bits are forwarded for each code bit. However, novel compression approaches based on the \ac{IBM} have been designed successfully to limit the information loss by quantization while keeping the relevant information \cite{b28}. It is expected, that this technique can significantly reduce the resulting \ac{FH} rate, while meeting the End-to-End performance constraints \cite{b29}. However, the application of \ac{IBM}-based quantization is beyond the scope of this paper.

The downsides are the considerably increased computation resources and power consumption's. Nonetheless, with the emergence of high-performance Artificial Intelligence-based solutions for channel estimation and equalization, this split shows promising potential, especially for non-terrestrial nodes.

\subsection{Option 6: Decoding}
This option involves several key steps: descrambling, rate dematching, and decoding, which prepare the data for transmission via the feeder link to the ground. Descrambling reorders the data to its original form, and rate dematching adjusts the data sequence to match the original transmission rate without altering the total number of bits. The decoding process, which is particularly crucial, takes \acs{LLR} bits, $Q_{\mathrm{LLR}}$, and reconstructs the original information bits. This process is performed using a specific decoding method (like \ac{LDPC} decoding), which operates at a code rate $R_{\mathrm{c}}$.  The decoding step reduces the number of remaining bits in the data stream, as it compresses the data by a factor related to the code rate. This is expressed as:
\begin{equation}
    R_{\mathrm{Opt6}} = R_{\mathrm{Opt7.3}} 
     \cdot \frac{1}{Q_\mathrm{LLR}} \cdot R_{\mathrm{c}} \;.
    \label{eq:fs6}
\end{equation}

Option 6 offers the advantage of markedly reducing the \ac{FH} data rate. However, the functionalities involved, particularly the decoding process, entail a considerable degree of computational complexity and energy consumption onboard.

\subsection{Option 2: High-Layer DU/CU split}

\ac{FS} Option 2, proposed by 3GPP as a higher layer split between the CU and DU at the output of \ac{RLC}, is recommended for regenerative scenarios, including \ac{NTN}, and can be considered a baseline because it is currently the only functional split specifically suggested by 3GPP for such applications. In this option, nearly all signal processing is performed on the platform, leading to significantly lower \ac{FH} requirements and more relaxed latency constraints, but at the cost of increased onboard complexity. The \ac{FH} data rate on the feeder link can be calculated similar to \cite{b8} as: 
 \begin{equation}
    R_{\mathrm{Opt2}} = PR\cdot \frac{B_{\mathrm{BW}}}{BW_{\mathrm{ref}}}\cdot \frac{N_{\mathrm{L}}}{N_{L_{\mathrm{ref}}}}\cdot \frac{M}{M_{\mathrm{ref}}} + signaling
 \end{equation}
 in which $N_{L_{\mathrm{ref}}} = 1$ as reference number of layer based on \cite{b10}. $signaling = N_{\mathrm{UE}}\cdot p_{\mathrm{UL}}\cdot S_{\mathrm{UL}}\cdot N_{\mathrm{L}}$, is the amount of data as function of  load on the system (as $p_{\mathrm{UL}}$ shows the percentage of users requesting \ac{UL}) and the average size of content in \ac{UL} by users. 

\section{Results}
\label{ch:results}

In this section we present the results of the analysis for the \ac{FH} data rate of each \ac{FS} option formulated in Section~\ref{ch:FH_BW_calc} for two described scenarios in Section~\ref{ch:model_and_formulation}.

\begin{figure*}[t]
\centering

\begin{minipage}{0.5\textwidth}
    \centering
    \scalebox{0.47}{\input{Figures/600km_10km_eMBB_S_and_Ka_Band.pgf}} 
    \caption{ \ac{eMBB} service type with $M = 16$ QAM and $R_{\mathrm{c}} = 0.64$.}{}
    \label{fig:result1_eMBB}
\end{minipage}%
\hfill
\begin{minipage}{0.5\textwidth}
    \centering
    \scalebox{0.47}{\input{Figures/600km_10km_mMTC_S_and_Ka_Band.pgf}} 
    \caption{\ac{mMTC} service type with $M = 4$ QAM and $R_{\mathrm{c}} = 0.66$.}{ }
    \label{fig:result1_mMTC}
\end{minipage}
\end{figure*}

Fig.~\ref{fig:result1_eMBB}, shows the needed \ac{FH} data rates for each functional split (FS) option for the two scenarios of \ac{LEO} satellite at $h_{\mathrm{C}} = \SI{600}{\km}$ and \ac{HIBS} at $h_{\mathrm{C}} = \SI{10}{\km}$ for the \ac{eMBB} service type in case of S and Ka bands, while Fig.~\ref{fig:result1_mMTC} shows the same for \ac{mMTC} service type. The platform altitude, service type, and $B_{\mathrm{BW}}$ directly influence the number of users that can be covered, which in turn affects the required \ac{FH} data rate on the feeder link.

The needed \ac{FH} data rate for a higher altitude node like \ac{LEO} satellite in SC1 is much higher than \ac{HIBS} in SC2, due to covering more cells and therefore users. For \ac{mMTC} results in Fig.~\ref{fig:result1_mMTC}. for both scenarios and $B_{\mathrm{BW}}$ the \ac{FS} options 8 to 6 all show much higher data rate in comparison to \ac{eMBB} in Fig.~\ref{fig:result1_eMBB} due to denser presence of users.

As anticipated, option 8 demands the highest data rate for all cases as time-domain IQ samples for all received signals, independent from the load on the system, are forwarded to ground-station on the feeder link. \ac{FS} options 7.1, 7.2 and 7.3 reduce the data rate in Fig.~\ref{fig:result1_eMBB} in average by $41.3\%$, $64.8\%$ and $89.7\%$, respectively, w.r.t option 8 by not adding much more extra complexity on the board. Option 6 hosts the decoder can reduce the rate for both frequency bands by $97.9\%$ and $83.34\%$ w.r.t option 8 and 7.3 respectively. The values are closely the same for \ac{mMTC} service type in Fig.~\ref{fig:result1_mMTC}. For option 2 as baseline we can see reduction of $99.46\%$ and $97.45\%$ w.r.t option 8 and 7.3 respectively, in cost of adding more functionalities on board. We can conclude that \ac{FS} option 6 can be a potential candidate  for non-terrestrial nodes, in comparison to option 2 with lower complexity and cost. Depends on the type of service \ac{FS} option 7.3 can come in the second place, mainly for \ac{eMBB}.

We can say that for \ac{mMTC} service option 2 shows the most applicable \ac{FH} data rate, while option 6 in S band also is feasible, especially since the beam diameter is wider than Ka band means more users can be covered.


In Fig.~\ref{fig:result2},  the required \ac{FH} data rates for \ac{FS} Options 7.3, 6, and 2 are shown as a function of different modulation orders ($M$) and corresponding coding rates ($R_{\mathrm{c}}$) for \ac{eMBB} services in scenario 2, where the \ac{HIBS} is positioned at $h_{\mathrm{C}} = \SI{10}{\km}$. The $R_{\mathrm{c}}$ for each $M$ is selected to maximize spectral efficiency, according to \cite{b30}. In case of air-born nodes such as \ac{HAPS} or drones, due to their lower altitude and longer visibility in comparison to \ac{LEO} satellites, the non-terrestrial node based on the request of ground station can ask users to change their \ac{MCS} based on the received signal \ac{SINR} to perform link adaptation on the service link. This leads to change in the \ac{FH} data rate on the feeder link. Option 7.3 increases more by higher modulation order due to the need of sending demodulated bits multiply by the $Q_{\mathrm{LLR}}$ factor. option 6 and 2 show close \ac{FH} data rate in lower ($M, R_{\mathrm{c}}$), while after 64 QAM the gap increases, due to the higher code rate. Option 7.3 with lower complexity in comparison to option 6 and 2, in both S band and Ka band, can be a suitable option even for higher modulation order especially in case of \ac{eMBB} service type for air-born nodes. This can be extended as a future work for case of other non-terrestrial nodes such as \ac{LEO} satellites by performing a link-level simulation.  
\begin{figure}[btp]
\centering
\scalebox{0.5}{\input{Figures/data_rates_vs_modulation_order.pgf}}
\caption{Scenario 2 needed \ac{FH} data rate of split options 7.3, 6 and 2 vs modulation order and Coding scheme pair for \ac{eMBB} service type. The X-Axis is in log-scale}
\label{fig:result2}
\end{figure}

\section{Conclusion}
\label{ch:conclusions}
This paper investigates the concept of \ac{FS} in the physical layer for components of 3D Networks. We have derived the required \ac{FH} data rate in \ac{UL} for non-terrestrial objects. Based on our results, we have shown that the data rates depend on key parameters, which we can optimized based on the specific requirements of a scenario, such as minimizing the \ac{FH} data rate versus increasing the computational complexity w.r.t the option 2 \ac{CU}-\ac{DU} split as baseline. Our formulation gives the flexibility to extend the analysis for different types of use cases as we analyzed the for different heights, two different type of service and frequency bands of S and Ka band. The selection of \ac{FS} option mainly depends on the use case specifically type of service and frequency band. As shown in case of \ac{eMBB} service type option 7.3 can be promising option, while for more users to be covered (higher $h_{\mathrm{C}}$ or service type of \ac{mMTC}) option 6 and option 2 are more feasible.

\vspace{12pt}

\end{document}